\documentstyle[epsfig]{europhys}

\begin{document}
\euro{70}{1}{29-35}{2005}
\Date{1 April 2005}

\shorttitle{Dissipative collapse of the adiabatic piston}

\title{Dissipative collapse of the adiabatic piston}

\author{R. Brito\inst{1}\footnote{Permanent address: Dpt.~F\'\i sica Aplicada I and GISC, Univ.~Complutense, 28040 Madrid, Spain},
M.J. Renne\inst{2} and C. Van den Broeck\inst{3}}
\institute{
\inst{1}{Institute Theoretical Physics, Universiteit Utrecht,
3508 TD Utrecht, The Netherlands}\\
\inst{2}{Burg.~v.d.~Heidelaan 55, 3451 ZS Vleuten, The Netherlands}\\
\inst{3}{Limburgs Universitair Centrum, B-3590 Diepenbeek, Belgium}}

\pacs{
\Pacs{05}{40.-a}{Fluctuation phenomena, random processes, and Brownian motion}
\Pacs{05}{70.Ln}{Nonequilibrium and irreversible thermodynamics}
\Pacs{45}{70.Mg}{Granular matter}
}

\maketitle

\begin{abstract}  An adiabatic piston,
separating two granular gases prepared in the same macroscopic state, is found to eventually collapse to
one of the sides. This new instability is explained by a simple macroscopic theory which is furthermore in
qualitative agreement with  hard disk molecular dynamics.
\end{abstract}

The problem of the so-called adiabatic piston was described in 1960 by  Callen \cite{callen}
and has received a lot of attention recently \cite{Lieb,AP}.
This construction consists of a cylinder containing two gases separated by an adiabatic piston.  It is of
interest because the principle of maximum entropy does not predict the equilibrium state
\cite{callen,Lieb}. In particular any state with gases left and right at equilibrium with the same pressure is
a possible equilibrium state of the compound system, even if the gases are at different temperatures. 
The degeneracy is however lifted when one moves away from the macroscopic limit. 
When fluctuations are taken into account, the set-up becomes a Brownian motor
\cite{bm} and the piston moves towards the cold chamber until the  
system relaxes to "full" equilibrium with equal temperatures and pressures in both gases. 
In this letter, we raise a different question: what happens when the gases are granular, i.e., they
gradually loose energy due to dissipative collisions. Our motivation is twofold. On the one hand, granular
matter has recently been the object of intensive theoretical and experimental research \cite{GM}. Somewhat
surprisingly, freely moving boundaries, which  can be easily realized in experiment,  have not received
much attention. On the other hand, we will show that the situation is in a sense more dramatic
than in the above mentioned non-dissipative case. Of particular interest to us is the case when both gases
are prepared in the same macroscopic state.  One expects that the piston will not move because of the
left-right symmetry, which is preserved as the gases cool on both sides by dissipative collisions.
However, our analysis will reveal that this state is unstable. Fluctuations or small disturbances will
induce a motion of the piston which amplifies and finally leads to a full collapse of one of the gases.
The origin of the instability is simple: the motion of the piston in a given direction will
compress the gas, increasing the energy dissipated by collisions and hence cool the gas,
leading to a further decrease of the pressure exerted from this side. This
simple intuition is confirmed by extensive molecular dynamics and a stability analysis of
macroscopic equations of motion.

To illustrate the type of behavior that is observed, we start by presenting some results from molecular
dynamics simulations for a two-dimensional system. We consider a rectangular container of
size $L_{x}\times L_{y}$, separated in two compartments by a piston of vertical length $L_{y}$ and mass $M$.
The latter can  slide without friction along the horizontal $x$-direction. Its position will be denoted by
$x \in [0,L_x]$. The compartments left and right, whose
corresponding variables will be identified by a subscript $i=1,2$, respectively, contain granular hard
disk gases  of
$N_{1}$ and $N_{2}$ particles, of diameter $\sigma $ and
mass $m$.  The disks move freely between collisions, while upon collision, total momentum is conserved but
 a fraction $1-\alpha ^{2}$ of their kinetic energy is dissipated, where
$\alpha$ is the\ so-called coefficient of normal restitution. The
piston itself does not conduct heat and undergoes
perfectly elastic collisions with the gas particles.

Figure 1 shows the typical behavior, starting from a  symmetric situation with the piston in the
middle, $x=L_x/2$, and equal densities and temperatures in left and right compartment. The parameter values
correspond to small dissipation,
$\alpha=0.99$, and initial low density, $n_1=n_2=0.04$. Clearly, the initial symmetry
is somehow broken leading to a final collapse of the piston to one side.  More details of the dynamics are
shown in fig.~2.a, including the piston position and temperature in both reservoirs versus time. The black dots
correspond to the configurations shown in fig.~1. At very short times, the piston performs
small-scale random deviations around the starting position, as a
result of the fluctuations in the pressure (fig.~2.a, inset).   As these deviations increase, the motion develops into more
regular oscillations combined with a net motion towards one of the sides (towards the left chamber in fig.1).  The piston
approaches one of the side walls while the gases cool down, and the oscillations shift to
lower frequency and amplitude.  In the final stage the piston completely
collapses, in a finite time, to one of the sides, with the particles reaching a close-packing structure, as
seen in the right panel of fig.~1.  From then on, the piston stops moving. In fig.~2.b, we reproduce the
results for higher dissipation, $\alpha=0.85$, with as most notable difference the complete suppression of
the oscillations. Note that the direction in which the piston moves is completely
determined by initial fluctuations. This is  confirmed by the fact that the collapse is, in different
realizations, equally often to each side, while the characteristic "onset" time for the observed
instability also changes from one realization to the other.

\begin{figure}
\label{fig1}
$$
\epsfig{file=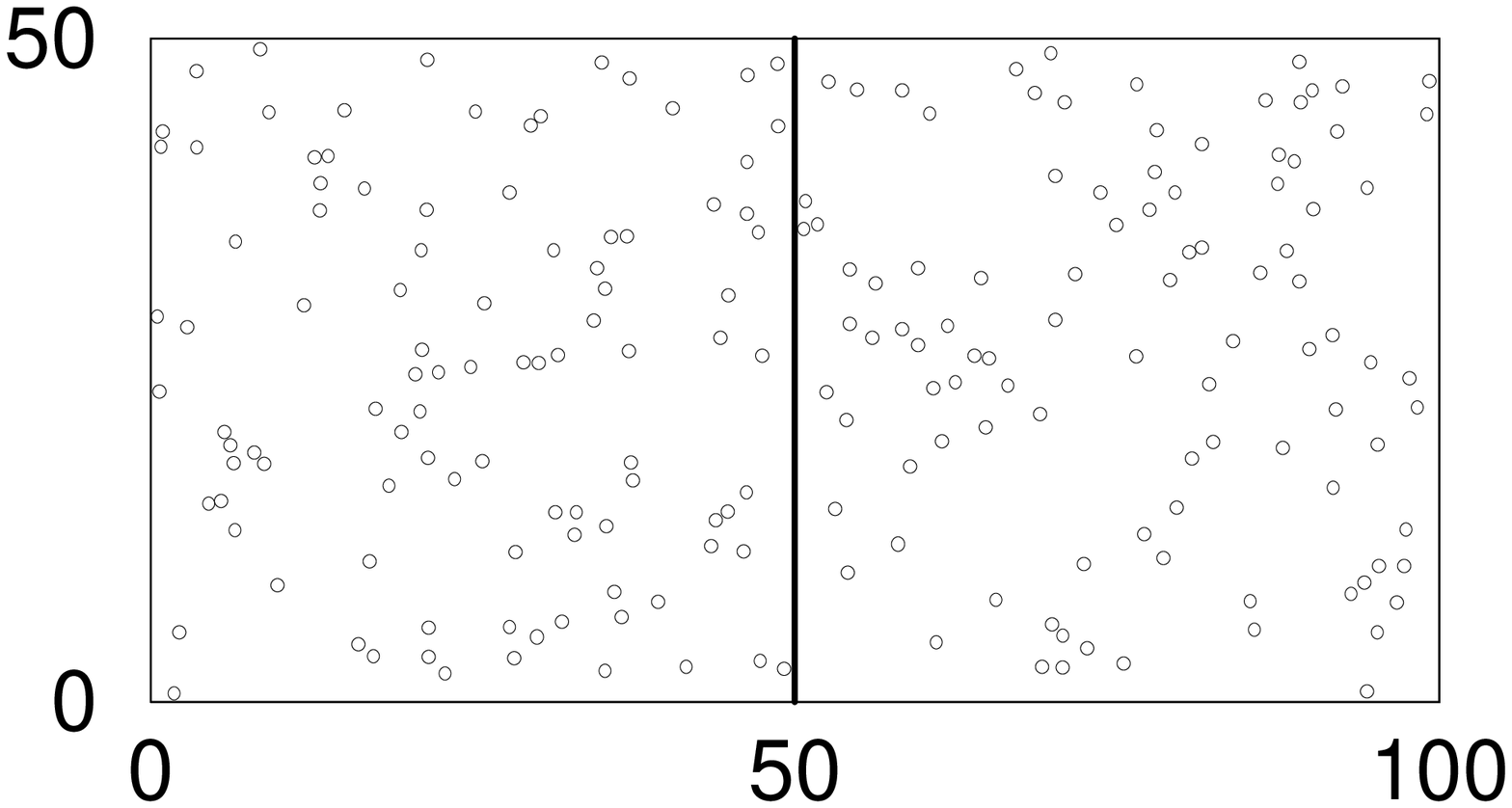,height=2.6cm}\epsfig{file=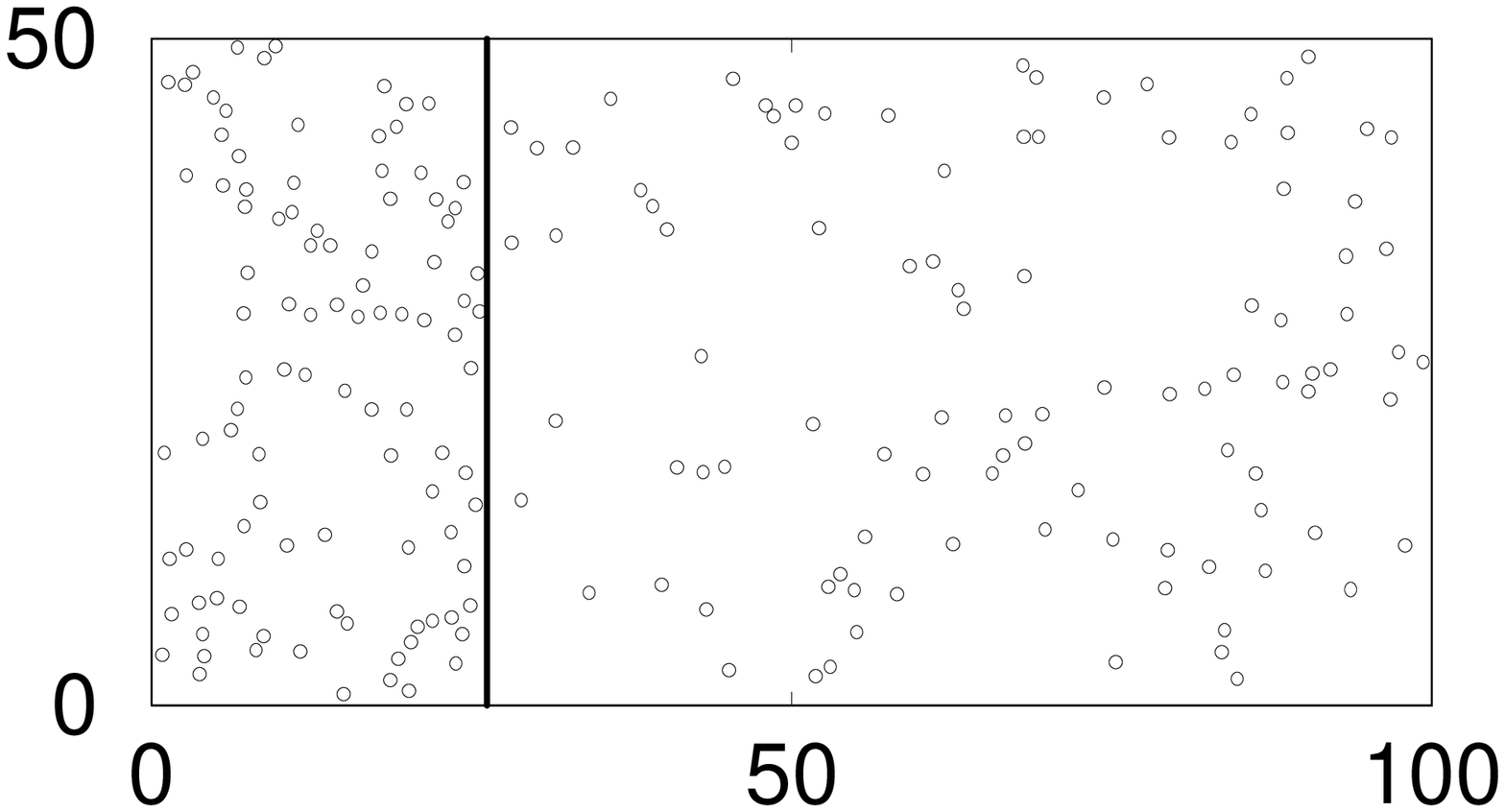,height=2.6cm}\epsfig{file=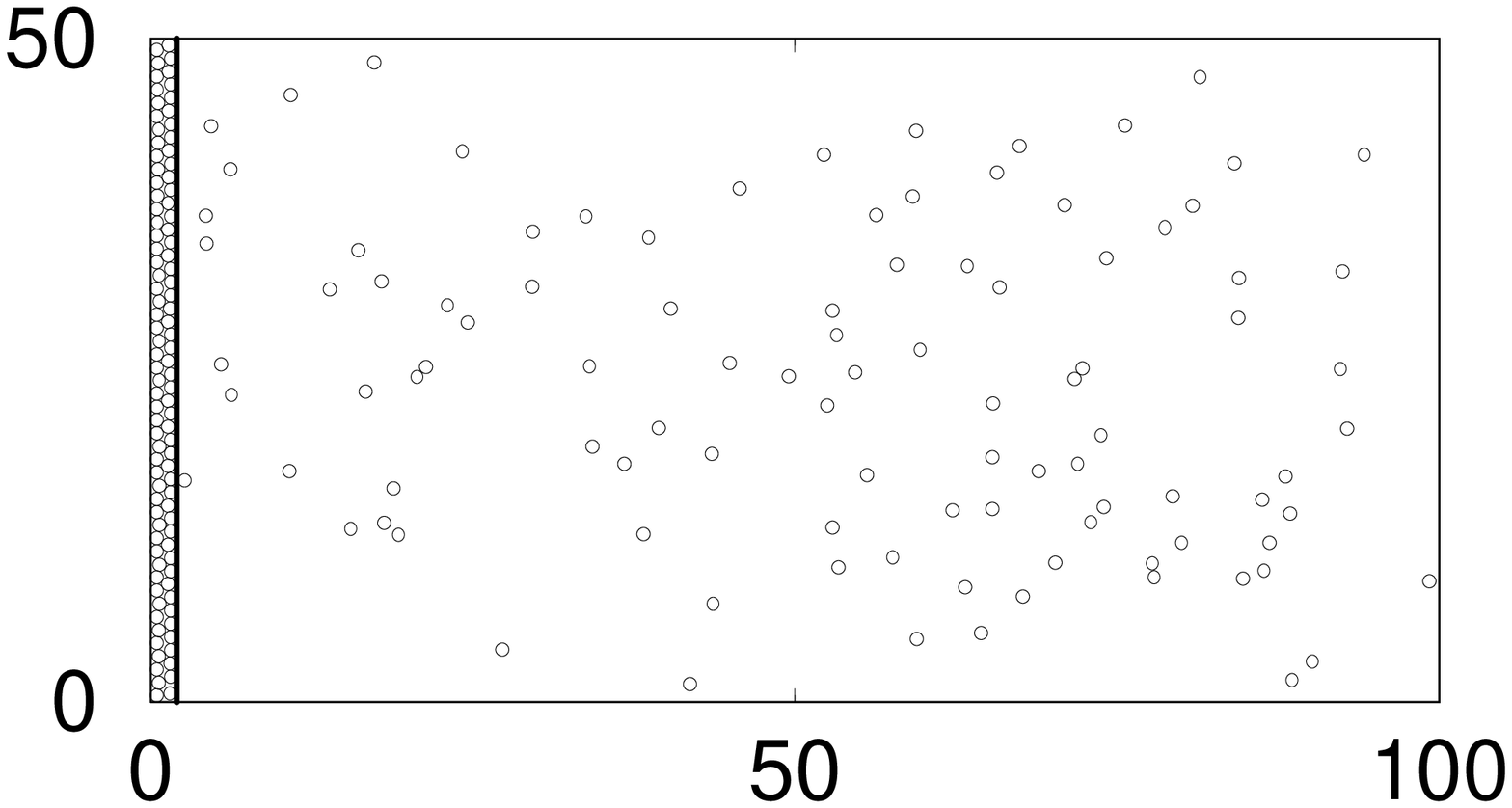,height=2.6cm}
$$
\caption{Time evolution of the system for the case of $N_1=N_2=100,\,T_1(0)=T_2(0)=1,\, \alpha=0.99,\, \sigma=1$ and
$M=2000m$, for $t=0$, $t=10^6$ and $t=10^7$. 
Piston dimensions are given in units of $\sigma$.
 Collapse of the piston to the left side is observed on the rightmost panel.}
\end{figure}

\begin{figure}
\label{fig2}
$$\epsfig{file=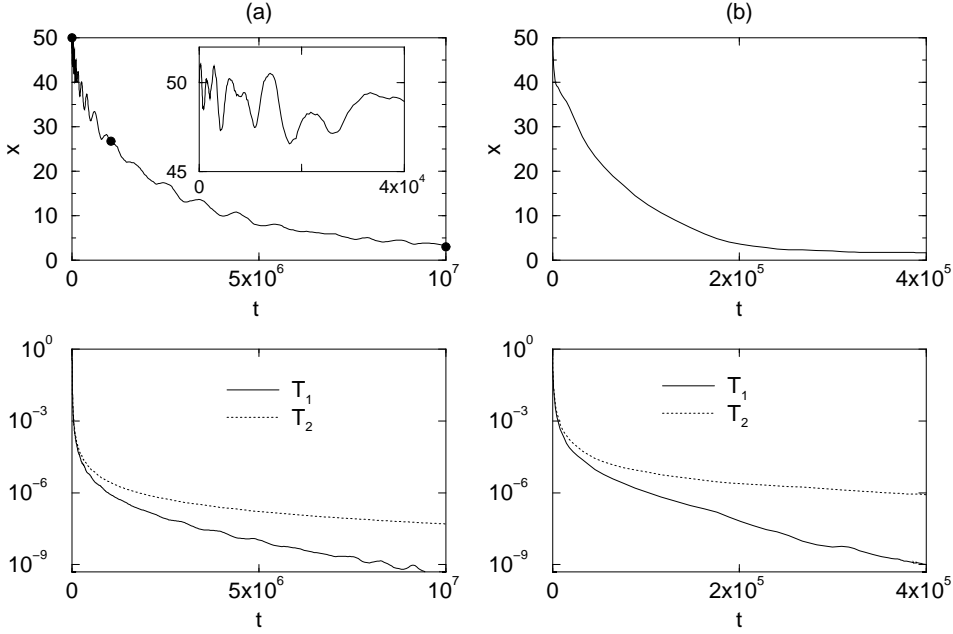,height=0.89\textwidth,angle=270}$$
\caption{Piston position and temperatures of the chambers versus time. Left column:
parameter set as in fig.~1.  Marked with black dots are the configurations plotted in fig.~1.
Right column, the same parameter set except $\alpha=0.85$.}
\end{figure}

To give a theoretical explanation of the observed behavior, we proceed with the derivation and analysis of macroscopic
equations of motion.
The energy in each chamber changes due to two mechanisms. The first one is compression or expansion
of the gas, and it is given by the expression  $dU_i=-P_idV_i$.
The second mechanism is the cooling due to inelastic collisions
that is described by Haff's law \cite{haff}, i.e., the amount of  energy $dU_i/dt|_C$
dissipated per unit time is proportional to the
number of collisions multiplied by the average energy lost per collision.
More precisely at low densities it has the form $ dU_i/dt|_C = A(1-\alpha ^{2})n_{i}k_B N_{i}T_{i}^{3/2}$.
Here $n_{i}\equiv N_{i}/V_{i}$ is the number density,
and $ A=\pi ^{1/2}$ $k_{B}^{1/2}\sigma m^{-1/2}$ a constant. The volumes are
$V_{1}=xL_{y}$ and $V_{2}=(L_{x}-x)L_{y}$.   We furthermore employ the equation $U_{i}=N_{i}k_{B}T_{i}$,
exact for hard disks, and assume, for simplicity, the ideal gas law for
the pressure, $ P_{i}={N_{i}k_{B}T_{i}}/{V_{i}}$.
Therefore, the variation of the temperature in each chamber is written as:
\begin{eqnarray}\label{Torig}
\frac{dT_{1}}{dt} &=& -\frac{T_1v}{x} -\frac{A(1-\alpha ^{2})N_1}{L_yx} T_1^{3/2}
\nonumber\\
\frac{dT_2}{dt}  &=& \frac{T_2v}{L_x-x} -\frac{A(1-\alpha ^{2})N_2}{L_y(L_x-x)} T_2^{3/2},
\end{eqnarray}
where $v$ is the speed of the piston, $v=dx/dt$.
Equations (\ref{Torig}) lead after simplification  to the following set of evolution equations for the temperatures:
\begin{eqnarray}\label{T}
-\gamma _{1}T_{1}^{3/2}  = \frac{d}{dt}xT_{1}, \qquad
-\gamma _{2}T_{2}^{3/2}  = \frac{d}{dt}(L_{x}-x)T_{2},
\end{eqnarray}
where the constants
$\gamma _{i}={A(1-\alpha ^{2})N_{i}}/L_{y}$
have been introduced. This set of equations is closed by the
equation of motion for the piston:
\begin{equation}\label{x}
M\frac{d^{2}x}{dt^{2}}=L_{y}(P_{1}-P_{2})=N_{1}k_{B}\frac{T_{1}}{x}%
-N_{2}k_{B}\frac{T_{2}}{L_{x}-x}.
\end{equation}

An analytic treatment of the above set of nonlinear
equations appears to be difficult. However, the question of stability of a (time-dependent) reference
state can be studied by analysis of the  linearized equations.
In particular the (in)stability of the symmetric state discussed in the above simulations can be investigated. More
precisely, considering
$N_1=N_2\equiv N$ (hence $\gamma_1=\gamma_2\equiv\gamma$), we note that
$x(t)=x(t=0)=L_{x}/2$ and $ T_{1}(t)=T_{2}(t)=T_{0}\left( 1+{\gamma t}{
}\sqrt{T_{0}}/L_{x}\right) ^{-2}\equiv T(t)$, where $T_0$ is the common initial temperature,
is an exact time dependent solution
of eqs. (\ref{T})-(\ref{x}) describing the symmetric cooling of
the system with fixed piston position. To study the stability of this state, we substitute
$x(t)=L_{x}/2+\xi (t)$ and $ T_{i}(t)=T(t)+\vartheta _{i}(t)$
into these equations, and linearize with
respect to $\xi (t)$ and $\vartheta _{i}(t)$ and their derivatives. We thus find:
\begin{eqnarray}\label{T1}
-\frac{3}{2}\gamma T(t)^{1/2}\vartheta _{i}(t)&=&\pm\frac{d}{dt}T(t)\xi (t)+\frac{%
1}{2}L_{x}\frac{d\vartheta _{i}}{dt},\\ \nonumber
\frac{ML_{x}}{2Nk_{B}}\frac{d^{2}\xi }{dt^{2}}&=&-\frac{4}{L_{x}}T(t)\xi
(t)+\vartheta _{1}(t)-\vartheta _{2}(t),
\end{eqnarray}
where the plus sign is for $i=1$ and minus sign for $i=2$. Integration of (\ref{T1})  leads to the following
explicit solution:
\begin{equation}\label{xt}
\vartheta _{1}(t)=-\vartheta _{2}(t)=-\frac{2}{L_{x}}T(t)\xi (t)+\frac{%
6\gamma }{L_{x}^{2}}T(t)^{3/2}\int_{0}^{t}dt^{\prime }\xi (t^{\prime }).
\end{equation}
corresponding to the relevant initial
conditions
$\vartheta _{1}(0)=\vartheta _{2}(0)=0$.
Substitution of these results in the eq. (\ref{T1})  gives an integro-differential equation for
$\xi(t)$.
By differentiating this equation with respect to time, and switching to a new time variable
$s \equiv 1+\gamma \sqrt{T_{0}}t/{L_{x}}$
one finds after some trivial rearrangements
the following third order differential equation:
\begin{equation}\label{xto}
2\Gamma s^{3}\frac{d^{3}\xi }{ds ^{3}}+6\Gamma s^{2}\frac{%
d^{2}\xi }{ds^{2}}+2s \frac{d\xi }{ds }-\xi =0,
\end{equation}
where $\Gamma={M\gamma^{2}}/({16Nk_{B}})={\pi(1-\alpha^{2}%
)^{2}nL_{x}M\sigma^{2}}/({32L_{y}}{m})$ and $n=2N/(L_{x}L_{y}).$ The new time
variable $s$ is proportional to $t$, and is not the time measured in units
of particle collisions as it is usually done in studies of granular fluids
\cite{GM}. Equation~(\ref{xto}) is homogeneous in $s$, and therefore its general
solution is given by: $\xi(s)=a_{1}s^{\mu_{1}}+a_{2}s^{\mu_{2}}%
+a_{3}s^{\mu_{3}},$ where the coefficients $a_{i}$ are determined by the
initial conditions (initial position, speed and acceleration) and where the
$\mu$'s are the roots of the equation $2\Gamma\mu^{3}+2(1-\Gamma)\mu-1=0$. For
$0<\Gamma<1$, this third order polynomial in $\mu$ has one positive real and
two complex conjugated roots. The positivity of the real root indicates the
announced instability of the reference state, while the presence of complex
conjugate roots show that the instability is accompanied by oscillations.

\begin{figure}[ptb]
\label{fig3}
\[
\epsfig{file=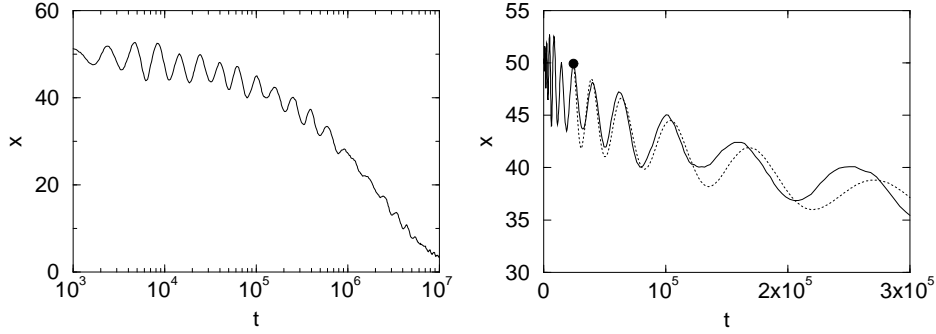,height=12.5cm,angle=270}%
\]
\caption{Left: Piston position for the parameter set of fig.~1 as a function
of $\ln(t)$ revealing oscillations in agreement with eq.~(7). Right:
Comparison of the simulation results with the analytic solution of eqs.~(1)
and (3), with initial condition corresponding to the dot marked in black.}%
\end{figure}

In the simulations, $\Gamma$ is actually very small: for instance, with
$\alpha=0.99$, $n=0.04,$ $M=2000m$ and $L_{y}=50,L_{x}=100$ (parameter set in
fig.~1), one finds $\Gamma\simeq6.2\times10^{-3}$. So for comparison, we can
use a small $\Gamma$ expansion for the roots: $\mu_{1}\simeq\frac{1}%
{2}[1+\frac{3}{4}\Gamma]$ and $\mu_{2}=\mu_{3}^{\ast}\simeq-\frac{1}%
{4}[1+\frac{3}{4}\Gamma]+\frac{i}{\sqrt{\Gamma}}[1-\frac{13}{16}\Gamma]$. As
initial conditions (remembering that $t=0$ is equivalent with $s=1$), we
consider the piston exactly in the middle, $\xi(s=1)=0$, with initial speed
$v$, ${d\xi}/ds(s=1)={L_{x}v}/(\gamma\sqrt{T_{0}})$, and zero
acceleration, ${d^{2}\xi}/ds^{2}(s=1)=0$. After evaluation of the
coefficients $a_{1},a_{2},a_{3}$, returning to the original time variable, and
introducing the time $t_{0}\equiv L_{x}/(\gamma\sqrt{T_{0}})$ and the phase
$\tan\phi\simeq2/(3\sqrt{\Gamma}),\,0<\phi<\pi/2$, we finally obtain:
\begin{eqnarray} \label{lineal}
\xi (t)\simeq \frac{3}{32} \frac{ A (1-\alpha^2)}{k_B}\frac{M v}{ \sqrt{T_{0}}}\frac{L_x}{L_y}
\left[ (1+t/t_0)^{1/2} -\frac{2}{3\sqrt{\Gamma }}(1+t/t_0)^{-1/4}\cos \{\phi +\frac{1}{\sqrt{\Gamma }}
\ln (1+t/t_0)\}\right] .
\end{eqnarray}
This expression for $\xi $, although derived in the linear approximation, reproduces
many of the features
of the simulation results for small dissipation: any initial speed,
however small, results in a superposition of a systematic motion with
oscillations of decreasing amplitude and increasing wavelength.
Quantitative agreement between theory and simulations can be achieved in the following way.
First, according to the above result, eq.~(\ref{lineal}), the oscillations of the piston position plotted
as a function of $\ln(t)$ have, for $t$ large enough, a constant frequency, proportional to
$1/\sqrt{\Gamma}$. Performing this plot from the numerical data, see fig.~3, we find
$\Gamma= 5.9\times 10^{-3}$, while the analytical value is $\Gamma=6.2\times 10^{-3}$.
From the same plot we can also extract the value of $\gamma$ by equating the difference
between the two maxima to $2\pi$. Here we obtain differences of the order of 3 to 9\% between analytic and numerical values.
Secondly, while we cannot fit the  solution (\ref{lineal}) directly to the simulations starting at time $t=0$, because the
initial phase there is dominated by the fluctuations, we can consider an
initial condition at a time $t$ after the random motion of the piston.
In fig.3, right panel, we have included the analytic solution  eq.~(\ref{lineal}) for the initial condition
marked as dot, and observe very good agreement with the simulations.

\begin{figure}
\label{fig4}
$$\epsfig{file=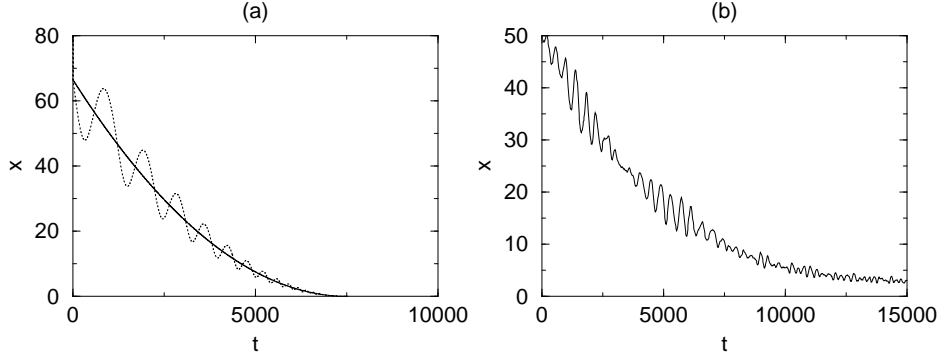,height=12.5cm,angle=270}$$
\caption{Left: Solution of eqs.~(2) and (3) where the pressure of chamber 2 is replaced by a constant value.
The solid line corresponds to the solution of eq.~(8), while the dotted line corresponds to a solution
which initial speed does not match eq.~(8). Right:
simulation results when the right chamber is replaced by a large one that provides an approximately constant pressure.
}
\end{figure}

The above linear theory is only valid for not too large times: the deviation
$\xi $ must be small, which means that the approximation is valid in the time regime $vt\ll
100\sigma \bar v/v$ where $\bar v\equiv (2k_BT_0/m)^{1/2}$. In particular it cannot describe the final relaxation of the piston.
We now proceed to show that  the set of differential equations (\ref{T}) and
(\ref{x}) predicts a final collapse of the piston. Let us assume that the piston is already far
displaced to one of the sides, say the left compartment, so that
$x\ll L_x$. Furthermore, we assume that the temperature in the right compartment does not change very much during the time
interval in which the collapse is taking place. Hence, the force  exerted by  the right compartment
remains approximately constant. With the replacement $F\simeq P_2 L_y$ a constant in (\ref{x}),  one verifies by
inspection that:
\begin{eqnarray}\label{col}
T_1(t)=T_1(0)\left(1-\frac{t}{t_c}\right)^2,\quad
x(t)=\frac{\gamma_1t_c}{4\sqrt{T_1(0)}}T_1(t),\quad t_c\equiv \frac{(8N_1k_B-M\gamma_1^2)\sqrt{T_1(0)}}{2\gamma_1F}
\end{eqnarray}
is an exact solution of (\ref{T}) and
(\ref{x}).
In words, both the piston position and the temperature in the collapsing compartment approach their final
values
$x=0$ and $T=0$ in a finite time $t_c$
following a parabolic trajectory.
This solution, eq. (\ref{col}), implies specific values for initial position and speed of the piston. The typical
solution, which we obtained by a numerical solution assuming a fixed force $F$
but with an initial speed that does not match eq.~(\ref{col}),
displays the same basic features but with
superimposed oscillations, see fig.~4.a. Furthermore the  frequency of these oscillations diverges as one approaches the final collapse.
Linearization around the solution  (\ref{col}) predicts a divergence as $1/\sqrt{x}$.
A quantitative comparison with simulations is difficult because the physical assumptions behind eqs. (\ref{T}) and
(\ref{x}) break down close to collapse. Nevertheless the basic features, collapse in a finite time
accompanied by oscillations of increasing frequency are clearly observed, see fig.~4.b.

We have focused the above discussion on the surprising properties of the symmetric case. We however also expect some
counter-intuitive behavior in the non-symmetric case. For example, consider as initial condition the
piston in the middle with equal temperatures left and right ($T_1(0)=T_2(0)=1$), but higher density (hence pressure) left
($n_1=0.02>n_2=0.008$). The time evolution of the piston position is shown in
fig.~5.  The initial displacement of the piston is as expected to the right side,
i.e., towards the compartment with low pressure. However, after some time the piston reverses its motion to end,
after a series of oscillations, collapsing to the left side. The surprising collapse of the high pressure
compartment appears to occur whenever the initial density difference is large enough. For the "complementary" case,
equal density ($n_1=n_2$), but higher temperature (hence pressure) left  ($T_1(0)>T_2(0)$),
one observes "normal" behavior with the piston eventually collapsing to the low pressure side.

To conclude, the macroscopic equations (\ref{T})-(\ref{x}), based on simple physical considerations,
capture surprisingly well the qualitative and quantitative features observed in the simulations.
Agreement between theory and simulations improves significantly when one includes finite size corrections \cite{gass}
both for
collision frequency in Haff's law and for the equation of state.
Other improvements will clearly require a much more involved theory
taking into account, e.g.,  the appearance of hydrodynamic flow and
inhomogeneities in density and temperature in the compartments, dissipative effects at the
boundary between piston and particles \cite{malek}, deviations from Haff's law \cite{BritoErnstEPL},
and thermal fluctuations. On the experimental side, we note that there is some similarity between the
phenomena observed here with the accumulation of granular gas initially distributed over several
compartments, into a single one \cite{Eggers}.

\begin{figure}
\label{fig5}
$$\epsfig{file=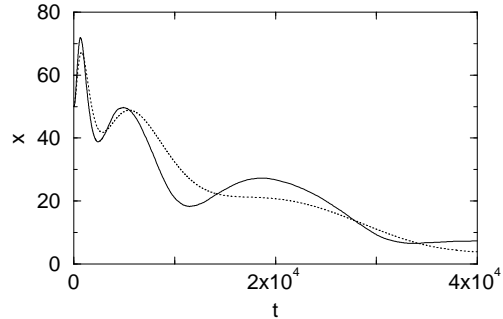,height=6.7cm,angle=270}$$
\caption{Piston motion for the case $N_1=50,\,N_2=20,\, T_1(0)=T_2(0)=1,\,L_x=100,\,L_y=50,\, M=2000m,\,\alpha=0.95$
resulting that  $P_1(0) > P_2(0)$, but eventually the piston moves towards the left chamber.
Solid line is the simulation result and dotted line is the solution of the macroscopic equations.}
\end{figure}

{\bf Acknowledgements.} The authors thank R. Kawai for useful discussions.
R.B. thanks S.~Redner for enlightening comments. 
R.B. is supported by Secretar\'\i a de Estado de Educaci\'on y Universidades (Spain)
and the research project FIS2004-271. M.J.R. acknowledges the
Dept. of Physics and Astronomy of the University of Utrecht for its hospitality.


\begin{thebibliography}{99}
\bibitem{callen}  H. B. Callen, "Thermodynamics" (Wiley, New York, 1960). 
\bibitem{Lieb} E.H. Lieb, Physica A, {\bf 263}, 491 (1999).
\bibitem{AP} C. Gruber, S. Pache and A. Lesne, J. Stat. Phys. {\bf 108}, 669 (2002). N.I. Chernov, J.L. Lebowitz, Ya.G. Sinai, Rus.
Math. Surv. 57, 1045 (2002). E. Kestemont, C. Van den Broeck
and M. Malek Mansour, Europhys. Lett. {\bf 49}, 143 (2000). M.J. Renne, M. Ruijgrok and Th.W. Ruijgrok,
Acta Physica Polonica B, {\bf 32} 4183 (2001).
\bibitem{bm} P. Reimann, Phys. Rep. {\bf 361}, 57 (2002).
\bibitem{GM} J. Duran, "Sands, Powders and Grains. An Introduction to the Physics of Granular Materials"
(Springer, New York, 2000). N. Brilliantov and  T. P\"oschel, "Kinetic Theory of Granular Gases" (OUP, Oxford, 2004).
\bibitem{haff}  P.K. Haff, J. Fluid Mech., {\bf 134}, 401 (1983). I. Goldhirsch and G. Zanetti, Phys. Rev. Lett. {\bf 70}, 1619 (1993).
\bibitem{gass} D. M. Gass, J. Chem. Phys. {\bf  54}, 1898 (1971); J. A. Barker and D. E. Henderson, Rev. Mod.
Phys. {\bf 48}, 587 (1976).
\bibitem{malek} M. Malek Mansour, C. Van den Broeck and E. Kestemont,  
 Europhys. Lett. {\bf 69}, 510 (2005).
\bibitem{BritoErnstEPL} R. Brito and M.H. Ernst, Europhys. Lett., {\bf 43}, 497 (1998).
\bibitem{Eggers} J. Eggers, Phys. Rev. Lett. {\bf 83}, 5322 (1999).
K. van der Weele, D. van der Meer, M. Versluis and  D. Lohse, Europhys. Lett. {\bf 53}, 328 (2001).
\end{thebibliography}
\end{document}